# Optical manipulation of shape-morphing elastomeric liquid crystal microparticles doped with gold nanocrystals

Yaoran Sun,[1,2] Julian S. Evans,[1] Taewoo Lee,[1] Bohdan Senyuk,[1] Patrick Keller,[3,4] Sailing He,[2,5] and Ivan I. Smalyukh[1,6,*]

[1]Department of Physics, Material Science and Engineering Program, Department of Electrical, Computer, & Energy Engineering and Liquid Crystal Materials Research Center, University of Colorado, Boulder, Colorado 80309, USA

[2]Centre for Optical and Electromagnetic Research, Zhejiang University, Hangzhou 310058, People's Republic of China

[3]Institut Curie, Centre de Recherche, CNRS UMR 168, Universite´ Pierre et Marie Curie, 75248 Paris cedex 05, France

[4]Department of Chemistry and Biochemistry and Liquid Crystal Materials Research Center, University of Colorado, Boulder, Colorado 80309, USA

[5]Department of Electromagnetic Engineering, Royal Institute of Technology, S-100 44 Stockholm, Sweden

[6]Renewable and Sustainable Energy Institute, National Renewable Energy Laboratory and University of Colorado, Boulder, Colorado 80309, USA

*Email: ivan.smalyukh@colorado.edu.

## Abstract

We demonstrate facile optical manipulation of shape of birefringent colloidal microparticles made from liquid crystal elastomers. Using soft lithography and polymerization, we fabricate elastomeric microcylinders with weakly undulating director oriented on average along their long axes. These particles are infiltrated with gold nanospheres acting as heat transducers that allow for an efficient localized transfer of heat from a focused infrared laser beam to a submicrometer region within a microparticle. Photothermal control of ordering in the liquid crystal elastomer using scanned beams allows for a robust control of colloidal particles, enabling both reversible and irreversible changes of shape. Possible applications include optomechanics, microfluidics, and reconfigurable colloidal composites with shape-dependent self-assembly.

Since its first demonstration, optical manipulation has always been an important research tool in physical sciences and biomedical research.[1–12] For example, in the study of soft matter systems, it allows one to explore inter-particle colloidal forces and provides insights into physical underpinnings of colloidal self-assembly.[10–12] Although laser trapping, alignment, rotation around different axes, and other types of non-contact optical control of both shape-anisotropic and birefringent particles have been demonstrated and widely used,[2–12] the efforts toward optical control of microparticle shapes are rare and the demonstrated shape transformations[8] are rather limited. For example, optical manipulation at relatively high laser powers ~100 mW allowed relatively modest stretching of red blood cells.[8] Liquid crystal elastomers (LCEs), on the other hand, have recently emerged as a new class of materials with facile control of structure and properties by means of thermal, electrical, and optical manipulation of orientational order through, for example, inducing transitions between nematic and isotropic states.[13–18] However, this shape manipulation is typically applied to centimeter-sized thin films rather than colloids. Control of elastomeric microparticles so far involved simple changes of the aspect ratio of anisotropic particles by means of thermally induced transformations between shapes of elastomeric particles pre-determined by the transition between nematic and isotropic states in the entire volume of LCE.[14–18] At the same time, dispersions of complex-shaped colloidal nano- and microparticles have recently attracted a great deal of research interest and may offer a means for fabrication of reconfigurable materials with desired optical and mechanical characteristics using shape-dictated self-assembly.[10–12] Methods for optical manipulation of colloidal shapes that would allow one to explore physical underpinnings behind such shape-dependent interactions are in a great demand.

In this work, we infiltrate cylindrical LCE microparticles with gold nanocrystals to enable efficient spatially localized energy transfer from a focused scanning laser beam into heat in a tiny submicrometer region of interest within the particle. Local heating alters the orientational ordering of the LCE molecules. By use of coupling between the shape and orientational ordering in LCEs, we demonstrate laser-induced reversible and irreversible transformation of the microcylinders to a number of complex-shaped colloids.

Micrometer-sized cylindrical actuators, made of side-on LCEs,[17] were obtained using replica molding, a soft lithography technique developed by Xia and Whitesides.[19] For a uniform alignment of the LCE director along the cylinder axis, particles were polymerized in a strong magnetic field of about 1 T oriented along the cylinder axis. The used LCE cylindrical particles were designed to have diameter (D) and length (L) dimensions of 2×4 µm, 17×70 µm, and 20×100 µm and were set free from a substrate by cutting them off using a razor blade. Dodecanethiol capped D = 2 nm gold nanocrystals in toluene were prepared a) using the biphasic Brust Schiffrin method.[20] The LCE microparticles were infiltrated with the gold nanocrystals in this toluene dispersion for 24 h. We replaced the solvent five times to remove non-infiltrated nanocrystals from the solvent while keeping them within the LCE. The LCE microparticles with thus embedded nanocrystals were then redispersed in water, silicone oil, or glycerol. In aqueous dispersions, surfactant sodium dodecyl sulfate (SDS, obtained from Aldrich) was added at 1 wt. % to improve colloidal stability of the fluid-borne LCE microparticles. The radius of the LCE particles slightly increases (by 5%–10%) while length decreases when they are dispersed in water, silicone oil, or glycerol as compared to original dimensions of the "dry" as-fabricated microcylinders (note that the effect of other nonpolar fluid hosts is very different and will be explored elsewhere).

We use a laser tweezers system consisting of a two-axis scanning-mirror head (XLRB2, Nutfield Technology) and a continuous wave Ytterbium-doped fiber laser (1064 nm, IPG Photonics).[21] The 1064 nm laser source was selected because this wavelength is about two times longer than that of the peak of absorption wavelength for gold nanoparticles (thus making the two-photon absorption processes efficient) and also because infrared light can penetrate deep into the bulk of LCE due to reduced scattering as compared to the visible spectral range. The tweezers setup enables steering of a focused beam along arbitrary computer-programmed trajectories and is integrated with a polarizing optical microscope (BX-51, Olympus). A sample is placed between two crossed polarizers and viewed in a transmission-mode polarizing optical microscopy (POM) while particles are manipulated via steering the beam. Control of linear polarization of the trapping beams is performed by means of rotating a λ/2-wave plate inserted immediately before the microscope objective. Three-dimensional (3D) director structures are probed using different modalities of nonlinear optical polarizing microscopy, including coherent anti-Stokes Raman scattering polarizing microscopy (CARS-PM), built around an inverted optical microscope (IX-81, Olympus) and described in details elsewhere.[22–24] In CARS-PM imaging, the target vibration modes are the aromatic C=C stretching at wavenumbers between 1400-1600 $cm^{-1}$,[16–18] because the average stretching direction is parallel to the long molecular axes of mesogenic units and yields polarization-dependent CARS-PM signals. Broadband stokes and 780 nm pump femtosecond pulses are used for excitation and the CARS-PM signal at $w_{\text{anti-Stokes}} = 2w_{\text{pump/probe}} - w_{\text{Stokes}}$ is detected with a band pass filter (BPF) having central wavelength at 700 nm and 13 nm bandwidth (1300–1600 cm$^{-1}$). We have also performed nonlinear optical imaging at the same excitation but without using a BPF, so that the images contain superimposed broadband CARS-PM signals from different chemical bonds and multiphoton excitation self-fluorescence of the LCE, as well as two-photon luminescence signals from gold nanoparticle aggregates.[25–27] For imaging and particle manipulation, we have used 10×–100× magnification microscope objectives with 0.1–1.4 numerical apertures. POM and CARS-PM (Figs. 1(a)–1(f)) reveal that the fluid-borne LCE particles have weakly undulating director field **n**(**r**) with the average director orientation along the cylinder's long axis (Figs. 1(g) and 1(h)). Observed undulations are likely caused by weak swelling of particles in the studied fluid hosts.

When heating in a fluid, we observe shrinking of the LCE particle's length and an increase of its diameter (Figs. 2(a)–2(c)). The most dramatic change of shape takes place in the vicinity of the phase transition of LCE from nematic to isotropic phase when the particles birefringence disappears as the LCE transitions to the isotropic state (Figs. 2(b) and 2(c)). The absorption of the 1064 nm laser light by the LCE is negligible, so that no substantial laser-induced heating and particle shape changes are observed at laser powers up to 2 W in the sample, unless the LCE is heated 1–2 degrees below the nematic-isotropic transition temperature. However, embedded gold nanoparticles allow for an efficient photothermal transfer of energy into heat and control of orientational ordering at laser powers less than 300 mW in the sample. When focusing the infrared laser beam into a point within the microcylinder, we observe that particle shape changes but then quickly relaxes back the original shape as the LCE cools down (Figs. 2(d)–2(h)), typically within about one second. In the region of a focused laser beam, the microparticle increases in width due to the local photothermal melting of the LCE (Figs. 2(d)–2(f)), consistent with experiments on heating of the entire sample, which also lead to the increase of cylinder width (Figs. 2(a)–2(c)). When the laser beam is scanned unidirectionally and along the cylinder axis, we observe that the particle "swims" in the direction opposite to the direction of scanning. Symmetric scanning of the beam back-and-forth does not cause such swimming motion but rather just oscillations of particle position.

Unidirectional nonreciprocal scanning of a laser beam along directions different from the microcylinder axis inevitably results in morphing of the particle shape, as demonstrated using the particle in a microchannel shown in Figs. 2(i) and 2(j). This unidirectionality is again essential to this effect, as bidirectional symmetric scanning along the same direction of the microparticle does not alter its shape. In all studied fluid hosts (Figs. 2(i) and 2(j) and 3), cylindrical particles bend toward the direction of scanning. Surprisingly, unlike in the case of local photothermal heating with a focused beam simply localized within the particle volume, the shape changes induced by nonreciprocal laser beam scanning persist after scanning is discontinued and can be stable for a long period of time (several months). This indicates that the observed scanning-induced shape transformations are caused by modifications of $\mathbf{n}(\mathbf{r})$ and polymer network of the LCE microcylinders. Nonreciprocal unidirectional scanning of the laser beam results in unidirectional motion of a molten region of the LCE within the microparticle, which subsequently causes reorganization of the polymeric network and $\mathbf{n}(\mathbf{r})$, triggering changes of shape that persist after the laser beam is turned off. This is confirmed by CARS-PM imaging of $\mathbf{n}(\mathbf{r})$ and chemical composition of elastomeric microparticles after the laser-induced shape modification (inset of Fig. 3(e)). LCE microparticles without gold nanospheres do not exhibit shape transformations in response to a scanned focused laser beam, unless they are heated to elevated temperatures of about one degree below the nematic-isotropic transition, in which case their response becomes reminiscent to that of gold-nanoparticle-doped LCE microcylinders at room temperature, although still less facile. Furthermore, these shape-morphing "hot" microparticles relax to the original cylindrical shape when laser scanning is discontinued, unlike in the case of nanoparticle-doped LCEs.

Although the used laser manipulation wavelength (1064 nm) is far from the wavelength of absorption peak of gold nanospheres (~530 nm), the photothermal laser heating is rather efficient. Nanoparticles in the LCE are likely forming small polydisperse aggregates, resulting in a sufficient light absorption at longer wavelengths, including that of the laser manipulation beam. On the other hand, two-photon absorption processes due to individual particles and their aggregates become important[25–27] at powers of the order of 100 mW of tightly focused infrared laser beams. Since a large fraction of light absorbed by metal nanoparticles is converted to heat[26,27] and local laser-induced elevation temperature of > 250 K/W can be achieved,[28] both of these processes contribute to the laser-induced local photothermal effect. Optical trapping of

LCE particles at powers of 20–30 mW in the sample can achieve robust spatial control due to the refractive index difference between particles and the host media as well as rotational control of the particles due to their birefringence and shape anisotropy. This can be accomplished even when particles are trapped in an array using multiple laser beams. For example, Fig. 4 shows how complex-shaped LCE particles can be aligned in different directions coinciding with adjustable beam's linear polarization directions controlled by a λ/2-wave plate (Fig. 4).

To conclude, we have demonstrated robust manipulation of the shape of colloidal birefringent elastomeric microparticles using infrared laser beams. In addition, these particles can be spatially translated, aligned, and rotated by means of polarized-light optical tweezers, simultaneously enabling all types of non-contact optical control. In addition to infrared laser beams, one can utilize laser sources with wavelength matched to the surface plasmon resonance peak. Furthermore, extension of this approach to azobenzene-based LCEs may allow for optical manipulation of microparticle shapes using substantially lower light intensities.[15] Possible applications include optomechanics, contact-free actuation of colloids, microfluidics, and reconfigurable colloidal composites with shape-dependent self-assembly.

This work was supported by the International Institute for Complex Adaptive Matter (YS) and by NSF Grants DMR-0820579 (TL), DMR-0844115 (BS), and DMR-0847782 (JSE, IIS). Work at Curie Institute was supported by ANR Grant ANR-2010-INTB-904-01 and by the CNRS-PICS "Curie-Tsinghua" project. We acknowledge technical assistance of P. Ackerman, Y. Lin, Z. Qi, and R. Trivedi and discussions with D. Broer, Q. Liu, and T. White.

**Figures**

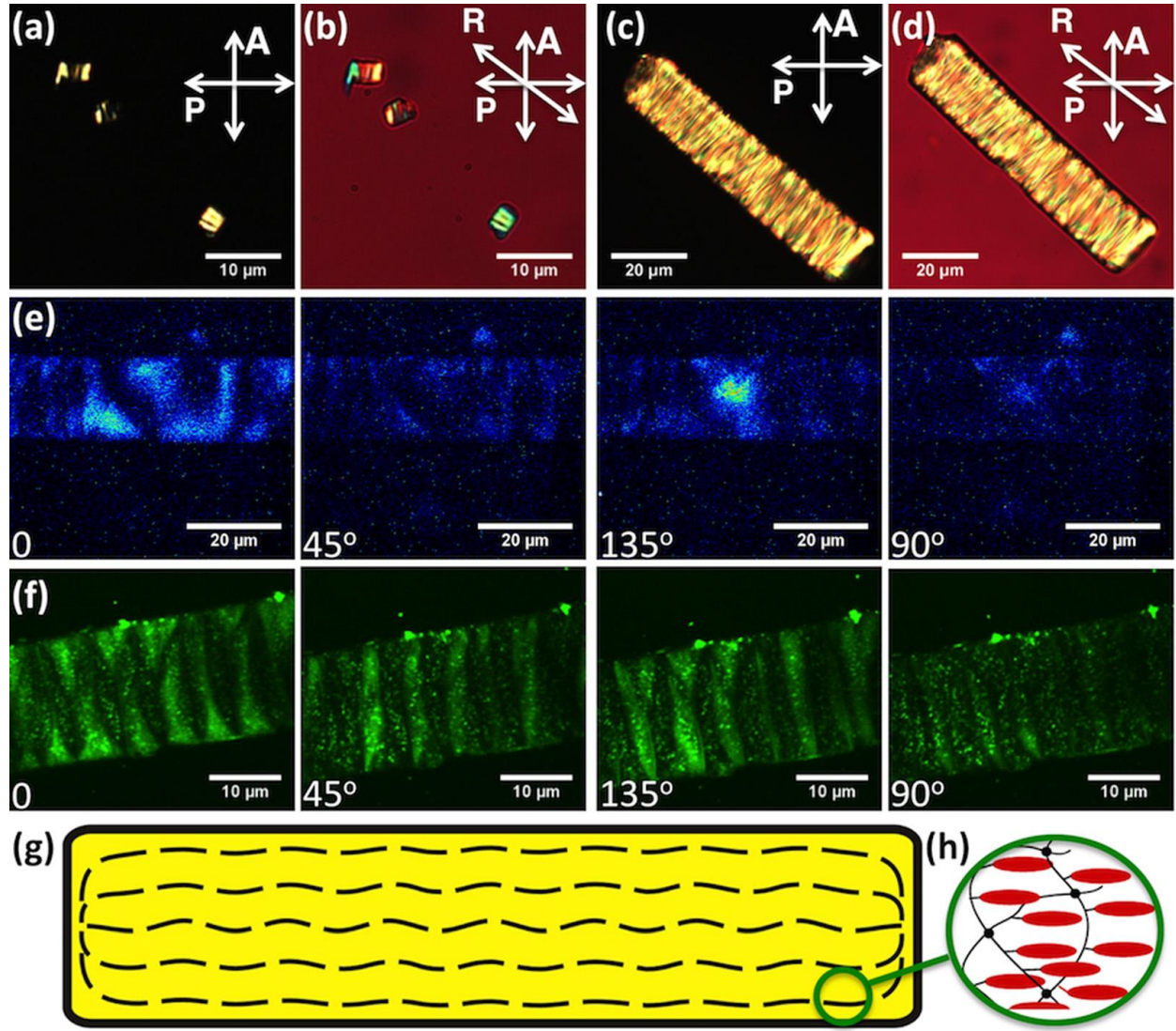

FIG. 1. Structure of cylindrical LCE microparticles. (a)-(d) POM images of LCE cylinders obtained (a) and (c) between crossed polarizer "P" and analyzer "A" and (b) and (d) with an additional full-wavelength (λ=530 nm) phase retardation plate with the "fast" axis marked "R." (e) and (f) Broadband nonlinear optical polarizing microscopy images obtained at different angles between the linear polarization of excitation laser beams and the cylinder axis in (e) a CARS-PM imaging mode and (f) with the same excitation as in CARS-PM but without a BPF. The angles between the cylinder axis and polarization of excitation light are marked in bottom left corners of images in (e) and (f). (g) and (h) Schematics showing (g) LCE microparticle's internal structure with weakly undulating **n(r)** (dashed lines) and (h) local alignment of cross-linked mesogenic units.

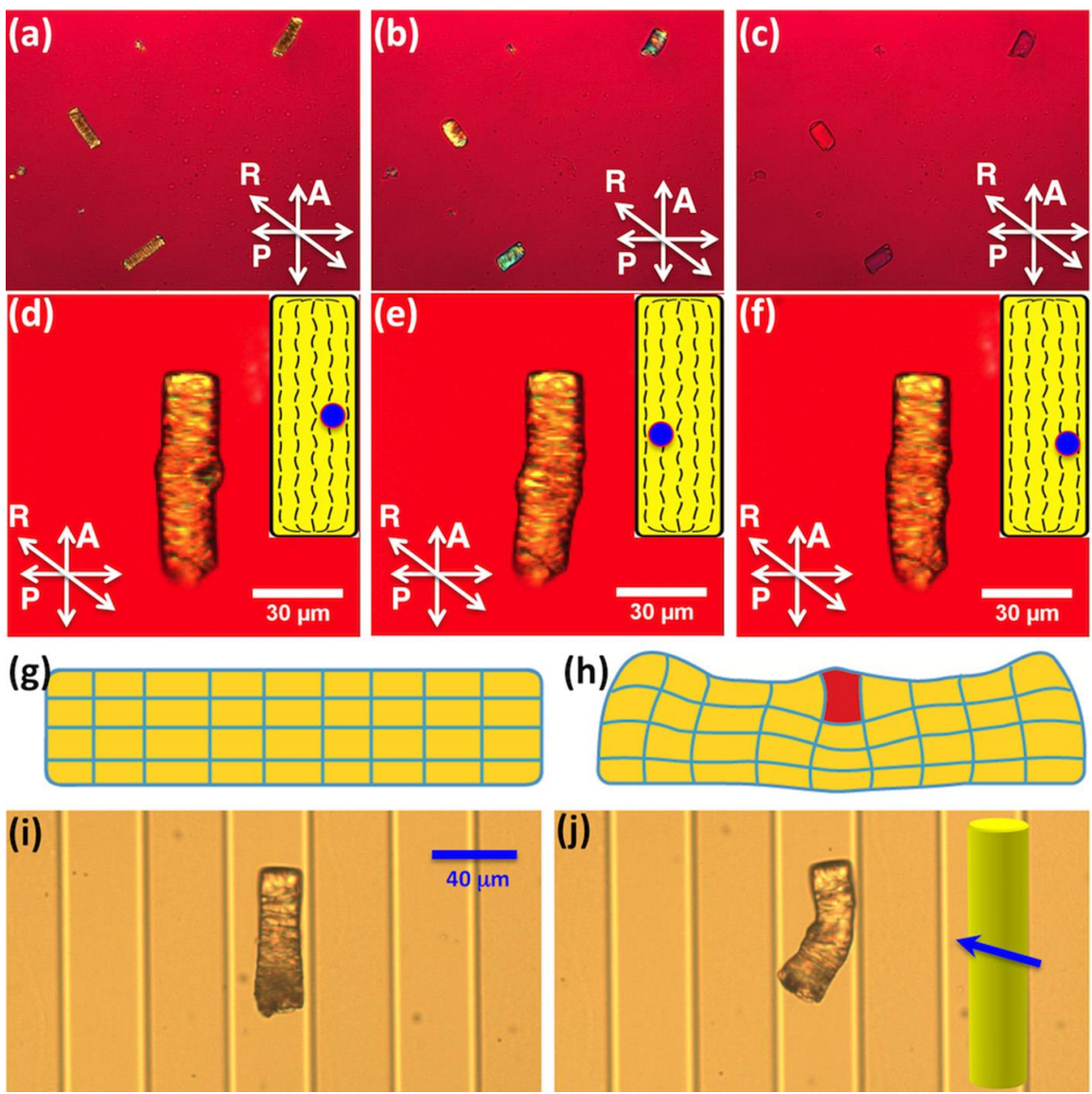

FIG. 2. Thermal and photothermal control of shape of LCE microparticles dispersed in glycerol. (a)-(f) POM images showing the change of the LCE cylinder aspect ratios as these particles are heated from (a) room temperature to (b) temperatures of about 1 °C below and (c) about 1 °C above the nematic-isotropic phase transition temperature of 135 °C. (d)-(f) Local manipulation of the microparticle shape by means of a tightly focused laser beam centered at locations shown by blue filled circles in the insets; note that these laser-induced changes of the microparticle shape are reversed after turning off the laser beam. (g) and (h) Schematics of initial (g) and laser-beam-modified (h) shape of the LCE microparticles deformed by local photothermal heating. (i) and (j) Reversible bending of the cylindrical microparticle suspended in glycerol inside a microchannel; this manipulation of the particle shape is achieved via repetitive unidirectional scanning of the focused laser beam along the direction of the blue arrow shown in the inset of (j).

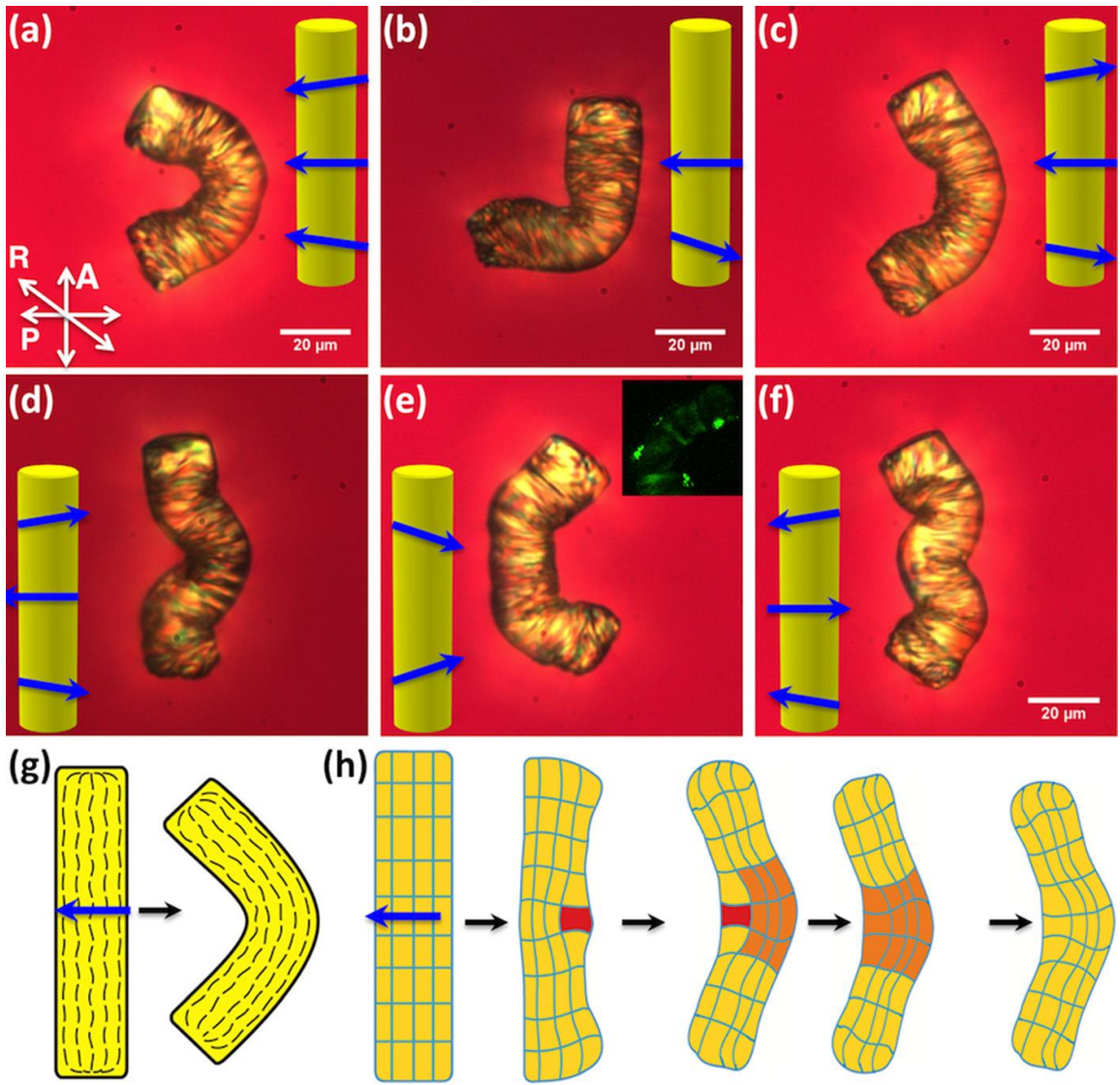

FIG. 3. Examples of robust reversible morphing of LCE microparticle shapes by means of unidirectional laser beam scanning along blue arrows shown in the insets of (a)-(f). The inset in (e) shows a typical CARS-PM image of a microparticle that was bent using a scanning laser beam. (g) Schematics of transformation of weakly undulating **n(r)** (shown by dashed lines) as the particle is bent due to scanning. (h) Schematic drawings show the effect of local manipulation of LCE orientational ordering by a scanned laser beam along the direction marked by the blue arrow; the scanning causes nonreciprocal unidirectional motion of a molten region (red) and in modification of the LCE polymeric network in the vicinity of the "hot" scanning region (orange), eventually leading to the stable modification of the particle shape that persists after the laser is turned off.

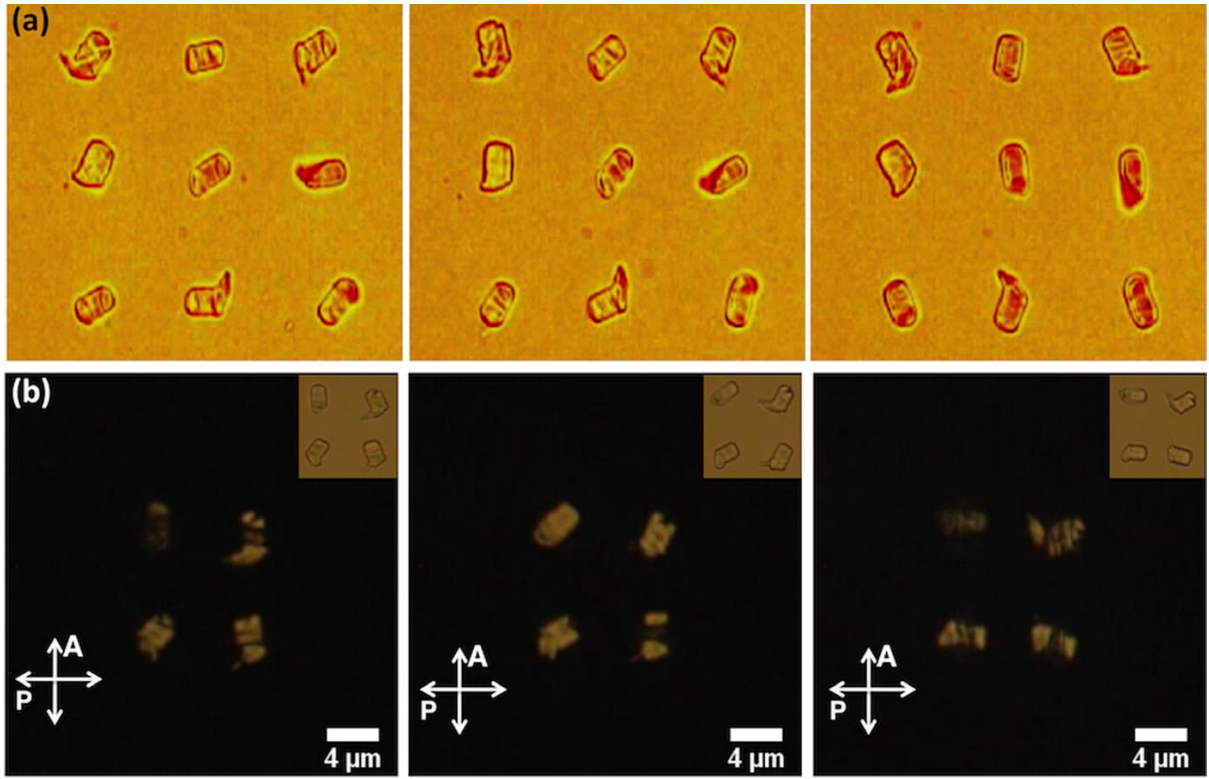

FIG. 4. Frames from bright-field (a) and POM (b) videos showing (a) alignment and (b) rotation of complex-shaped birefringent LCE colloids in trap arrays produced by means of holographic optical tweezers. Note that due to imperfections of colloidal shapes of arrays of trapped particles shown in (a) and (b), they align slightly differently with respect to the linear polarization direction of the trapping beam in (a) and rotate (b).